\documentclass[twocolumn,aps,prl,preprintnumbers,superscriptaddress]{revtex4-2}
\usepackage[utf8]{inputenc}
\usepackage{mathrsfs}
\usepackage{amsmath}
\usepackage{amsfonts}
\usepackage{amssymb}
\usepackage{amsthm}
\usepackage{multirow}
\usepackage{color}
\usepackage{graphicx}
\usepackage{pifont}
\usepackage{lettrine}
\graphicspath{{./images/}}
\usepackage{geometry}
\usepackage{hyperref}
\usepackage{lineno}
\usepackage{textcomp}

\hypersetup{
     unicode=false,          
     pdftoolbar=true,        
     pdfmenubar=true,        
     pdffitwindow=false,     
     pdfstartview={FitH},    
     pdftitle={My title},    
     pdfauthor={Author},     
     pdfsubject={Subject},   
     pdfcreator={Creator},   
     pdfproducer={Producer}, 
     pdfkeywords={keyword1} {key2} {key3}, 
     pdfnewwindow=true,      
     colorlinks=true,       
     linkcolor=blue,
     citecolor=blue,        
     filecolor=blue,      
     urlcolor=blue           
}
\usepackage[toc,page]{appendix}
\usepackage{ulem}

\begin{document}
\makeatletter
\@ifundefined{textcolor}{}
{
\definecolor{BLACK}{gray}{0}
\definecolor{WHITE}{gray}{1}
\definecolor{RED}{rgb}{1,0,0}
\definecolor{GRAY}{rgb}{0.4,0.4,0.4}
\definecolor{GREEN}{rgb}{0,1,0}
\definecolor{BLUE}{rgb}{0,0,1}
\definecolor{CYAN}{cmyk}{1,0,0,0}
\definecolor{MAGENTA}{cmyk}{0,1,0,0}
\definecolor{YELLOW}{cmyk}{0,0,1,0}
}
\makeatother

\title{Polarization transfer force on ferroelectric domain walls}
\author{Huanhuan Yang}
\affiliation{WPI-AIMR, Tohoku University, 2-1-1 Katahira, Sendai 980-8577, Japan}
\affiliation{Center for Spintronics Research Network, Tohoku University, Sendai 980-8577, Japan}
\author{Peng Yan}
\affiliation{School of Physics and State Key Laboratory of Electronic Thin Films and Integrated Devices,
University of Electronic Science and Technology of China, Chengdu 610054, China}
\author{Gerrit E. W. Bauer}
\email{bauer.gerrit.ernst.wilhelm.d8@tohoku.ac.jp}
\affiliation{WPI-AIMR, Tohoku University, 2-1-1 Katahira, Sendai 980-8577, Japan}
\affiliation{Center for Spintronics Research Network, Tohoku University, Sendai 980-8577, Japan}
\affiliation{Institute for Materials Research, Tohoku University, 2-1-1 Katahira, Sendai 980-8577, Japan}

\begin{abstract}
We investigate the dynamics of ferroelectric textures driven by polarization currents. We show that, ferrons, the quanta of collective polarization excitations, provide an exotic driving mechanism for domain wall (DW) dynamics, compared with their magnonic counterparts. By mapping the linear polarization dynamics of a DW onto a Schrödinger-like problem with a P\"{o}schl-Teller potential, we show that polarization waves are fully transmitted and therefore do not exert a net force on the DW in the linear regime. However, intrinsic nonlinearities give rise to a negative radiation pressure that pulls the DW toward the source. This mechanism  allows efficient DW control by optical excitation and temperature gradients with application potential in ferroelectric memory and logic devices.
\end{abstract}
\maketitle  

{\it  \color{blue} Introduction.} 
Ferroelectric domain walls (DWs) are topological defects central to ferroelectric functionalities by mitigating depolarizing fields and facilitating polarization switching \cite{Tagantsev2009}. While most DWs are believed to be of the Ising type, theory and experiments revealed Bloch \cite{Huang1997,Hlinka2011,He2024}, Néel \cite{Wei2016}, charged (head-to-head or tail-to-tail) \cite{Gureev2011}, and mixed types \cite{Lee2009,Gu2014} across multiple ferroelectric families \cite{Nataf2020,Seidelbook}.  Depending on the polarization profile, DWs  may be electrically neutral,  carry positive or negative bound charges, or a charge dipole \cite{He2024}.  The DW dynamics and its control is central for efficient polarization switching and therefore the design of ferroelectric nanoscale memory and logic devices \cite{Whyte2015,Sharma2017,Jiang2018,Bednyakov2018}.  

In magnets, DWs can be propelled by magnetic fields, spin-polarized electron  \cite{Tatara2008} or magnon \cite{Yan2011,Han2019} currents, and gradients in the magnetic anisotropy or temperature \cite{Hinzke2011}. Ferroelectric DWs have traditionally been actuated by applying electric fields \cite{Shin2007} and mechanical stresses \cite{Sumigawa2020}. 
Here we address a polarization current-based strategy to control ferroelectric textures. Closely analogous with magnons, ferrons are the quanta of polarization waves as predicted \cite{Bauer2021,Tang2022ferron,Tang2022,Adachi2023}, and observed  \cite{Wooten2023,Choe2025,Shen2025,Zhang2025} in various ferroelectric materials. However, a direct translation of the spin-transfer-torque  in magnetism to a polarization-transfer-force in ferroelectrics is not possible since in contrast to magnon currents, ferron currents are in general not conserved, even in the absence of dissipation. Nevertheless, in this Letter we report that ferron currents can drive ferroelectric DWs, but by a completely different physical mechanism.

We derive the Landau-Khalatnikov-Tani (LKT) equations of motion for Ising wall driven by a ferron current. Linearization around a static DW with a typical tanh profile maps the problem onto the Schrödinger-like equation with a reflectionless Pöschl-Teller potential, so in linear response ferron waves exert no force. However, nonlinear effects generate a momentum surplus behind the DW, giving rise to a negative radiation pressure that pulls the DW toward the ferron source. We quantify this mechanism by direct numerical simulations of the LKT equation, establishing polarization transfer forces on ferroelectric DWs as the counterpart of the spin transfer torque on magnetic DWs.

{\it \color{blue}Ferroelectric DW profile}. We consider a planar ferroelectric texture $p(x)$ in the $\hat{x}$ direction and a Landau-Ginzburg free energy density 
\begin{equation}\small
\begin{aligned}
{\mathcal F} =  f_0 + \frac{1}{2}\alpha p^2 + \frac{1}{4}\beta p^4+\frac{1}{2}g|\partial_x p|^2,
\end{aligned}
\end{equation}
where $\alpha<0$ for $T<T_c$ and $\beta$ are Landau coefficients, $g$ is the stiffness.

The DW profile
\begin{equation}\label{DWProfile}
p=-p_{s}\tanh [(x-X)/\Delta],
\end{equation}
minimizes the free energy and corresponds to the standard neutral Ising wall.
Here, $X$ is the wall center coordinate and $p_s=\sqrt{|\alpha|/\beta}$ is the spontaneous polarization amplitude, while the corresponding DW width is
$\Delta=\sqrt{2g/|\alpha|}$.

\begin{figure}
  \centering
  \includegraphics[width=0.48\textwidth]{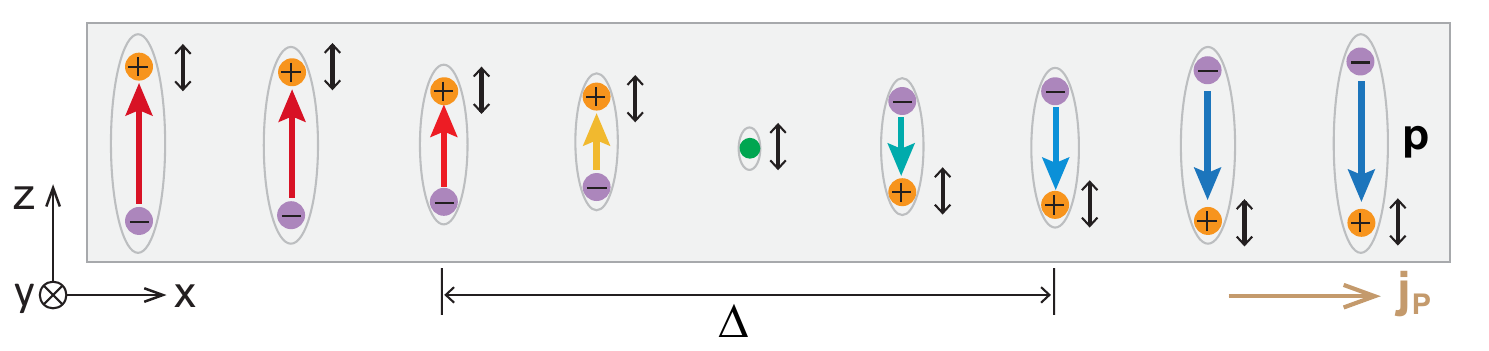}
  \caption{Illustration of a transverse and therefore neutral Ising wall with width $\Delta$ and polarization distribution $p$ that can be driven by a ferron current ${\bf j}_p$.}\label{F2}
\end{figure}

{\it \color{blue} Polarization dynamics.}
Here we introduce an unconventional method to drive DWs by a ferron current as illustrated by Fig. \ref{F2}. The LKT equation for a DW
\begin{equation}\label{LKT}
m_p \ddot{p} + \gamma \dot{p} - g \partial_x^2 p + \alpha p + \beta p^3 =0
\end{equation}
governs the spatiotemporal polarization dynamics, where $m_p$ is the polarization inertia and $\gamma$ is the damping coefficient \cite{Tang2022ferron}. 
We isolate the slowly moving rigid kink solution $p_0(x,t)$ from the total polarization dynamics  and expand the remainder ferron dynamics
$\tilde p(x,t)=p(x,t)-p_0(x,t)$ by
\begin{equation}
\begin{aligned}
\tilde p(x,t)\simeq &~\tilde p_0(x)
+\frac{1}{2}\tilde p_1(x)e^{-i\omega t}  
+\frac{1}{2}\tilde p_1^*(x)e^{i\omega t}\\
&+\frac{1}{2}\tilde p_2(x)e^{-i2\omega t}+\frac{1}{2}\tilde p_2^*(x)e^{i2\omega t}+\cdots
\end{aligned}
\end{equation}
to second harmonics of the drive frequency \(\omega\). Averaging Eq. \eqref{LKT} over the fast oscillations yields
\begin{equation}\label{LKTave}
m_p \ddot{p}_0+ \gamma \dot{p}_0+\mathcal{L}\tilde p_0
+ 3\beta p_0\left\langle \tilde p^{2}\right\rangle = 0
\end{equation}
where $\mathcal{L}:= \alpha+3\beta\,p_0^2-g\partial_x^2$  and  $\langle \tilde p^2\rangle$ is the mean-square average of the polarization fluctuations over the fast time scale.

We focus on the leading term in the expansion into collective coordinates, \textit{viz}.  the center $X(t)$ of a rigid DW texture
$p_0(x,t)\approx p_0[x - X(t)]$. Multiplying both sides of the Eq. \eqref{LKTave} by $\partial_x p_0= -\partial_X p_0$
and spatially integrate,  $\mathcal{L}\tilde p_0$ vanishes and we arrive at Newton's equation of motion of a damped oscillator
\begin{equation}\label{CCE}
m_p I \ddot{X} + \gamma I \dot{X} = F,
\end{equation}
where the overlap integral $I = \int dx(\partial_x p_0)^2=4p_s^2/(3\Delta)$
renormalizes the optical mass \(m_p\) and the force
\begin{equation}
F = 3\beta \int dx\, p_0(x)\,\partial_x p_0(x)\,\big\langle \tilde p^{2}(x)\big\rangle.
\end{equation}
Since $p_0(x)\,\partial_x p_0(x)$ changes sign at the DW center, $F$ is finite only when $\langle \tilde p^2\rangle$  on the sides of the DW differs. We now show that ferron current causes such an imbalance and exerts a ``radiation pressure" on the DW at rest.

{\it  \color{blue}Linear solutions}. 
We treat the dynamics as small fluctuations around the static kink solution $p_0(x)=p_s\tanh(x/\Delta)$ induced by an incoming plane polarization wave with amplitude \(A\). We then expand the polarization as
\begin{equation}
\begin{aligned}
p(x,t)=p_0+A \delta p^{(1)}
+ A^{2}\, \delta p^{(2)}+ \mathcal{O}(A^{3}).
\end{aligned}
\end{equation}
Substituting this into Eq.~\eqref{LKT} and collecting terms to first order in $A$,
\begin{equation}\label{LKT1o}
m_p \ddot{\delta p}^{(1)}
+ \gamma \dot{\delta p}^{(1)}
+\mathcal{L} \delta p^{(1)}
= 0.
\end{equation}

\begin{figure}
  \centering
  \includegraphics[width=0.48\textwidth]{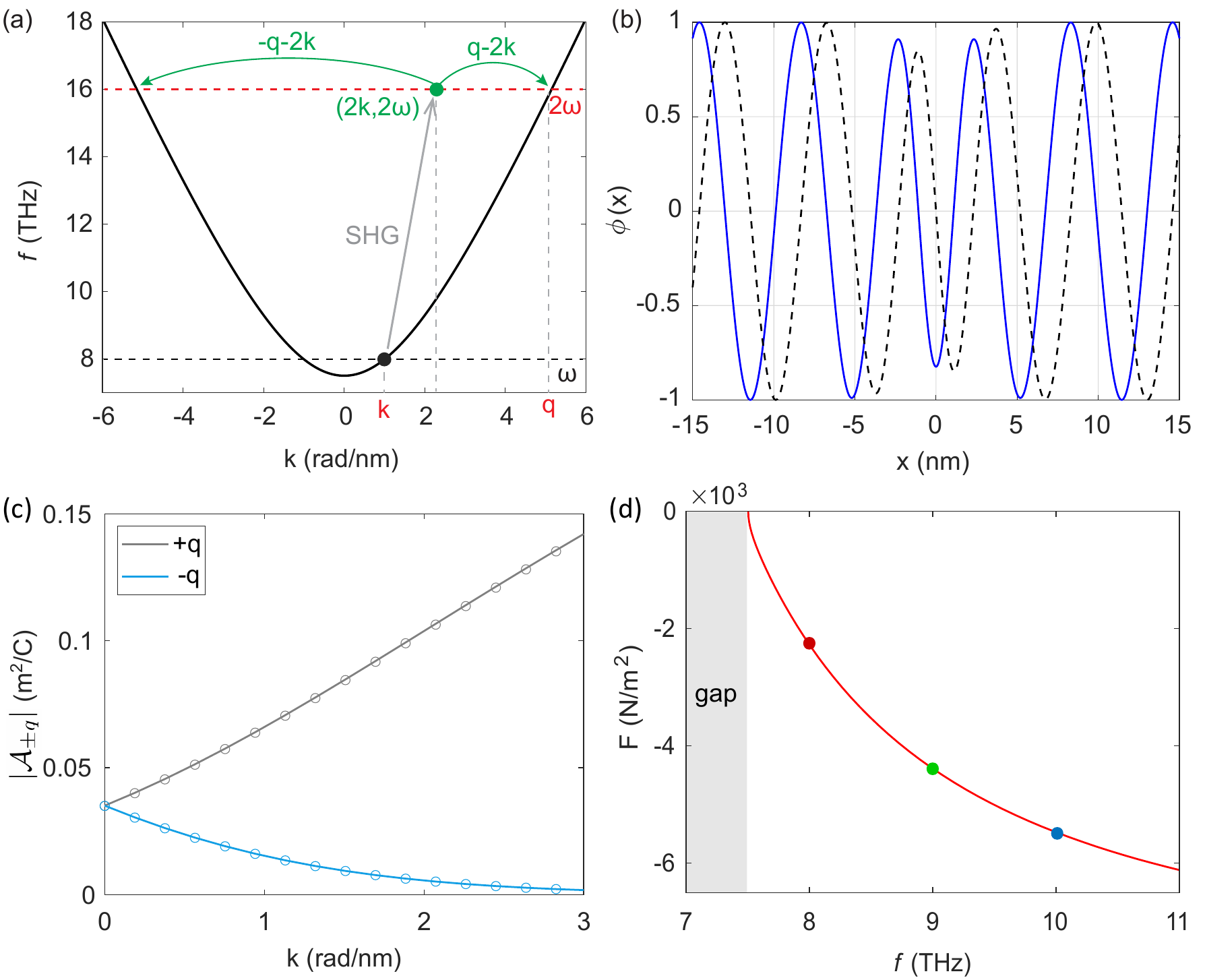}
  \caption{(a) Dispersion relation of ferrons for the material parameters of LiNbO$_3$ \cite{Para}. The gray arrow indicates the second-harmonic generation (SHG) in the sample, while the green arrows denote their energy-conserving but momentum-non-conserving scattering  $(2k,2\omega)\rightarrow(\pm q,2\omega)$ by the DW. (b) Wavefunction of a propagation ferron at $k=1$ rad/nm and frequency \(\omega\) [black dot in (a)]. The solid and dashed curves correspond to the real and imaginary parts, respectively. (c) Amplitude of the forward [$+q$ (gray)] and backward [$-q$ (blue)] scattering states of the second-harmonic components. (d) Radiation pressure $F$ as a function of the excitation frequency  $f=\omega / (2 \pi)$ and $A=0.1 p_s$. Dots mark the force at frequency $f=8,9,10$ THz, respectively.}\label{F3}
\end{figure}

The eigen problem for the harmonic ansatz $\delta p^{(1)}=\frac{1}{2}\phi_k(x)e^{-i\omega t}+\frac{1}{2}\phi_{-k}(x)e^{i\omega t}$ and \(\gamma=0\) can be mapped on a Schrödinger-like equation with an {\it index-2} reflectionless P\"{o}schl-Teller potential \cite{PTpotential}
\begin{equation}\label{EigenEq}
\left[- \frac{\partial ^2}{\partial x^2} - \frac{6}{\Delta^2}{\rm sech}^2\frac{x}{\Delta} \right]\phi_k(x) = \epsilon_k\phi_k(x),
\end{equation}
where $\epsilon_k=m_p\omega^2/g - 4/\Delta^2$ is energy of the eigenstate with index \(k\). 
This equation can be solved exactly \cite{Flugge1994}. There are two bound states  with $\epsilon_k<0$ that describe the vibration and breathing of the texture \cite{Chen2021}.
The continuum of propagating solution with $\epsilon_k>0$ become plane waves with wave number \(k\) far from the DW. Their eigenvalues and wave functions are  
\begin{equation}
\begin{aligned}
&\epsilon_k =k^2,~\omega_k=m_p^{-1/2}(gk^2-2\alpha)^{1/2};\\
&\phi_k(x) =
\frac{3\tanh^2\!\dfrac{x}{\Delta} - 1 - (k\Delta)^2
      - 3ik\Delta\tanh\!\dfrac{x}{\Delta}}
       {\sqrt{\bigl[(k\Delta)^2 + 1\bigr]\bigl[(k\Delta)^2 + 4\bigr]}}\,
  \exp\!\left( i k x \right).
\end{aligned}
\end{equation}
Since states with $k>0$ and $k<0$ do not mix, there is no reflection. Figure \ref{F3}(a) shows the dispersion relation of the propagating states for a LiNbO$_3$ ferron gap of $7.5$ THz \cite{Para}. The corresponding wave function in Fig. \ref{F3}(b) experiences only a $k$-dependent phase shift
\begin{equation}\label{Phase}
\Delta\varphi= -2\arctan\left[\frac{3k\Delta}{2-(k\Delta)^2}\right](\mathrm{mod}~2\pi)
\end{equation}
when passing through the DW region. The force $F$ vanishes because $\delta p^{(1)}$ has the same amplitude in the left and right domains. Deviations from the tanh profile that generate a P\"{o}schl-Teller potential with a non-integer index, could cause reflection and linear momentum transfer, but should be very small.

Spin waves move magnetic DWs even in the absence of backscattering because angular momentum conservation implies its transfer to the DW under transmission. However, we show in the supplementary material \cite{SM} that in ferroelectric textures the linear momentum is conserved and not the polarization. In linear response and disregarding dipolar and polaritonic effects,  the static dipole moment of a ferron changes sign when traversing the DW without feeling or exerting a force. However, we show in the following that the situation changes qualitatively once nonlinearities are taken into account.

{\it \color{blue}Nonlinear scattering.}
In the absence of damping the leading nonlinear term \(\delta p^{(2)}\propto A^{2}\) solves
\begin{equation}\label{LKT2o}
m_p \ddot{\delta p}^{(2)}
+ \mathcal{L}\,\delta p^{(2)} = f^{(2)},
\end{equation}
with source term
\begin{equation}
\begin{aligned}
f^{(2)} &= -3\beta p_0\big(A\delta p^{(1)}\big)^2 \\
&= A^2\!\left[f^{(2)}_{0}(x) + f^{(2)}_{2\omega}(x)e^{-2i\omega t}
+ f^{(2)}_{-2\omega}(x)e^{2i\omega t}\right],
\end{aligned}
\end{equation}
where $f^{(2)}_0 = -\frac{3}{4}\beta p_{0} \phi_{k}\phi_{-k}$,
$f^{(2)}_{2\omega} = -\frac{3}{4}\beta p_{0} \phi_{k}^{2}$, and
$f^{(2)}_{-2\omega} = -\frac{3}{4}\beta p_{0} \phi_{-k}^{2}$.
\(f^{(2)}_0\) generates a static contribution $\delta p^{(2)}_0(x\to\pm\infty)
=\pm 3/(8p_s)$. This is the finite ferron dipole that does not vanish in the quantum limit and is the analog of the finite angular momentum of a magnon \cite{Bauer2021,Tang2022ferron}. The last two terms introduce the solution $\delta p^{(2)}(x,t)=\delta p_{2\omega}^{(2)}(x)e^{-2i\omega t}+\delta p_{-2\omega}^{(2)}(x)e^{2i\omega t}$. In the absence of a DW, the second harmonic polarization is the plane wave $\delta p^{(2)}_{\mathrm{hom},2 \omega}(x)=-e^{2ikx}\beta p_s/(8\alpha)$.

Turning to the scattering processes in the presence of a DW, the dc ferron dipole switches sign but does not generate a momentum imbalance or force on the DW, viz. electric polarization is not conserved. However, the second harmonics at $(2k,2\omega)$ interact with the symmetry-breaking DW. Assuming dominant elastic scattering, they decay into the waves $(\pm q,2\omega)$ with amplitudes \({\mathcal A}_{\pm q}\), as sketched in Fig.~\ref{F3}(a), where $q=\sqrt{(4m_p\omega^2+2\alpha)/g}$ is the ferron wave number at frequency $2\omega$. 
In the far field, the general solution is the sum of a source-driven particular part and the solutions of the homogeneous differential equation \cite{Romanczukiewicz2004,Forgacs2008}
\begin{equation}
\delta p^{(2)}_{q}(x\to\pm\infty)
= -\frac{\beta}{8\alpha}\,p_s^\pm\,\phi_k^2
+ \mathcal{A}_{\pm q}\,\phi_{\pm q}(x).
\end{equation}
The nonlinear scattering amplitudes can be obtained using a Green’s function approach
$\mathcal{A}_{\pm q} = -\int dx\,\phi_{\mp q}(x)\, f^{(2)}_{2\omega}(x)/(gW),$
where $W=\phi'_{q}\phi_{-q}-\phi_{q}\phi'_{-q}=-2iq$ is the Wronskian. Evaluating the integral yields the closed-form result
{\small
\begin{equation}
\mathcal{A}_{\pm q}
= -\frac{3\pi}{2p_s}
\frac{(k\Delta)^{2}+4}{(k\Delta)^{2}+1}
\sqrt{\frac{(k\Delta)^{2}+4}{(q\Delta)^{2}+1}}
\frac{1}{q\Delta\sinh\left[ (2k\mp q)\Delta\pi/2 \right]},
\end{equation}}%
\noindent which is plotted in Fig. \ref{F3}(c) as a function of wave number.

The scattering does not conserve momentum, which implies linear momentum transfer and a force on the scattering object that is proportional to the incoming momentum flux $\Pi=\frac{|\alpha|}{64p_s^2}A^4k^2\Delta^2$ at $(2k,2\omega)$ \cite{SM}. The forward- and backward-propagating components with $q-2k$ and $-(q+2k)$ momenta, respectively, exert a radiation pressure
\begin{equation}\label{Force}
F = \frac{|\alpha|}{4} A^4 q\Delta^2\Big[(2k-q)|{\mathcal A}_{+q}|^2+(2k+q)|{\mathcal A}_{-q}|^2\Big].
\end{equation}
Substituting \(F\) into Newton’s equation \eqref{CCE}, we arrive at the steady-state DW velocity 
\begin{equation}\label{dotX}
v=\dot{X}=F/(\gamma I).
\end{equation}
In Fig.~\ref{F3}(d) we invert \(\omega(k)\) of Fig.~\ref{F3}(a) to obtain the force as a function of frequency.

\begin{figure}
  \centering
  \includegraphics[width=0.48\textwidth]{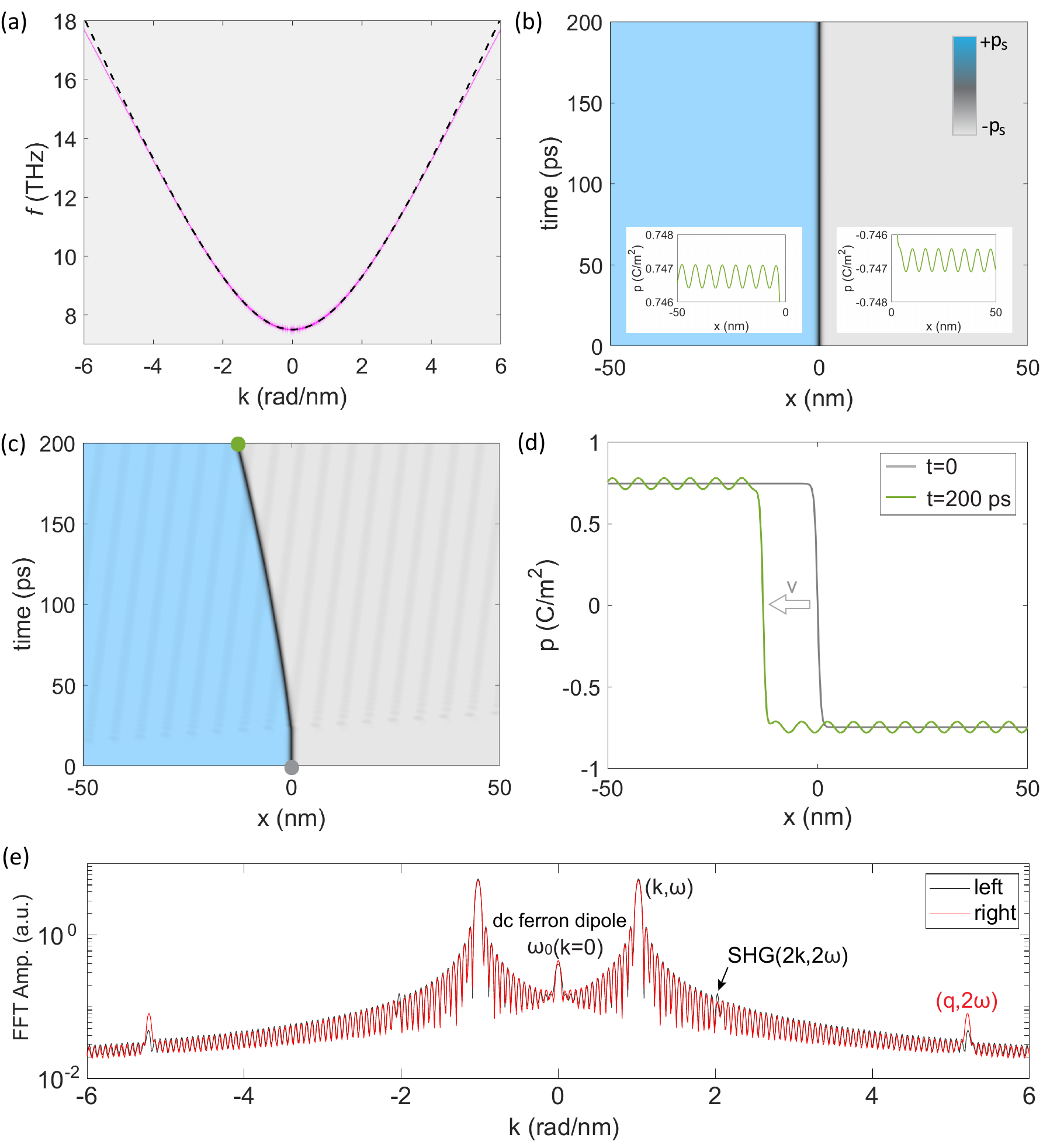}
  \caption{(a) Ferron dispersion in single-domain LiNbO$_3$. The color map shows the spectrum obtained by Fourier transforming the numerical results of $p(x,t)$ under broadband excitation and compares with the analytical dispersion (dashed line). (b) Spatiotemporal map of the polarization under a weak harmonic drive with frequency $f=8$~THz and amplitude $e_0=0.001$ applied at $x=-100$ nm. Insets: snapshots of the polarization dynamics to the left and right of the DW at $t=200$~ps after switching on the drive. The DW does not move in the linear regime. (c) Spatiotemporal map of the polarization under strong pumping $e_0=0.1$. (d) Polarization profiles at $t=0$ (gray) and $t=200$~ps (green). (e) Fourier amplitude of polarization fluctuations in the left (black) and right (red) domains, averaged over $t=150\sim200$ ps. In the simulations, $k=1$ rad/nm and $q=5.14$ rad/nm.}\label{F4}
\end{figure}

{\it \color{blue}Simulations}. We validate our theory by numerically solving the LKT equation \eqref{LKT} for a chain of $N=2001$ dipoles. We first consider a uniform domain and excite ferrons at $x=0$ by a broadband field pulse ${\bf e}(t)=e_0 \operatorname{sinc}(2\pi f t)\hat z$ with $f=\omega/(2\pi)=20$ THz and dimensionless electric field $e_0=0.1$ (or physical field $E_0=|\alpha|p_se_0=1.5$ MV/cm for LiNbO$_3$). We Fourier transform the computed time series $(x,t)$ into the frequency domain $(k,\omega)$ to obtain the ferron spectrum, shown in Fig. \ref{F4}(a) as a color map. The analytical bulk ferron dispersion
$\omega(k)={m_p}^{-1/2}(gk^2-2\alpha)^{1/2}$ agrees very well with the numerical results.

Next, we introduce an Ising wall profile centered at $x=0$ and launch ferrons by a \textit{weak} sinusoidal electric field ${\bf e}(t)=e_0\sin(2\pi f t)\hat z$ with $f=8$ THz and $e_0=0.001$ at the position $x=-100$ nm. As shown in the insets of Fig. \ref{F4}(b), the ferron current  is fully transmitted through the wall, consistent with the predicted reflectionless scattering.  Momentum is conserved, ferrons exert no force on the DW and the wall does not move.

Non-linear effects emerge when increasing the driving amplitude to $e_0=0.1$. According to Fig. \ref{F4}(c) the DW moves towards the source of the excitation  at $x=-100$ nm. After switching on the drive, the DW accelerates up to a constant velocity of \(\sim\)100 m/s. Figure \ref{F4}(d) displays the polarization profile at $t=0$ and $t=200$ ps [gray and green dots in (c)].  The Fourier transforms of the steady-state signals in the left (black) and right (red) domains in Fig. \ref{F4}(e) reveal the origin of the DW dynamics. While the peak at $k=0$ represents the static dipole carried by each ferron, the enhanced $2\omega$ signal on the right side is a consequence of a strong forward scattering of the second-harmonic content found in Fig. \ref{F3}(c). The momentum surplus in front of the kink exerts a net force that pushes the wall toward the incident wave, i.e., a negative radiation pressure.

\begin{figure}
  \centering
  \includegraphics[width=0.49\textwidth]{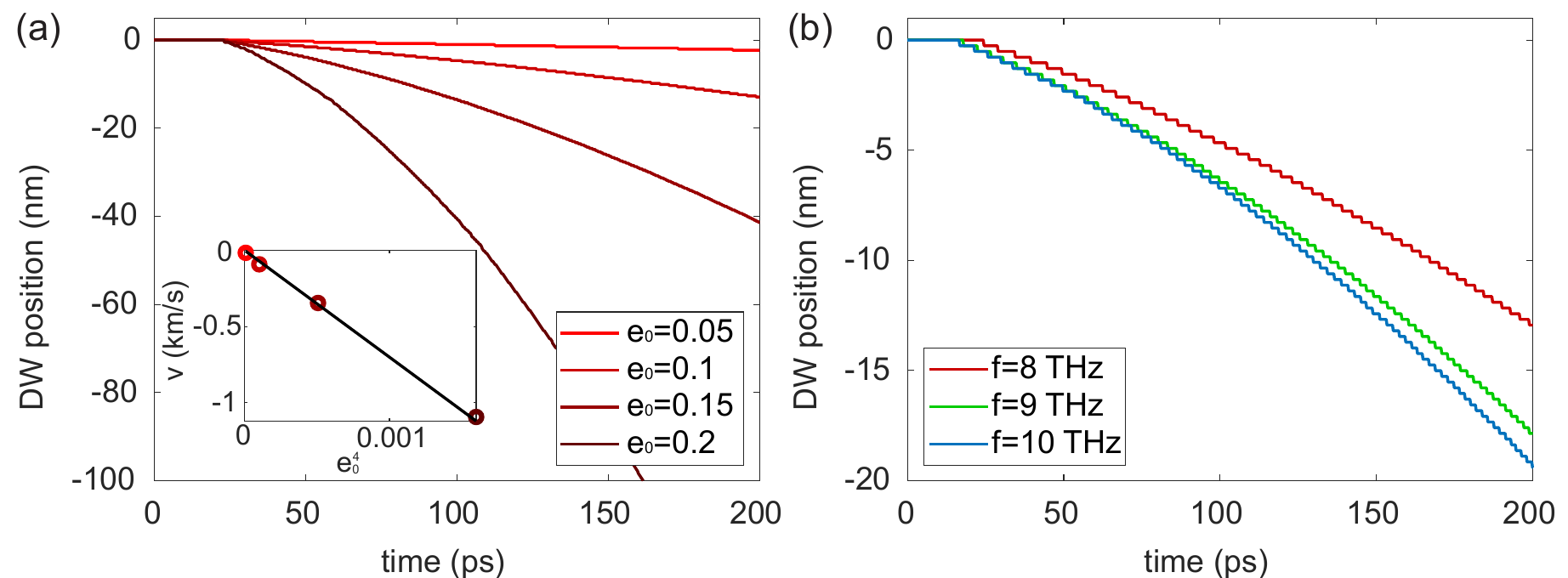}
  \caption{(a) Time-dependent DW position for several driving field amplitudes at 
$f=8$ THz. Inset: DW velocity as a function of the fourth power of the ferron amplitude. The numerical (dots) and analytical [Eq. \eqref{dotX}, solid lines] results agree well. (b) DW trajectories as a function of time for different driving frequencies at a fixed amplitude $e_0=0.1$.}\label{F5}
\end{figure}

This effect persists over a broad range of driving frequencies and amplitudes. Figure \ref{F5} summarizes the DW motion under different drives. Figure \ref{F5}(a) shows that increasing the electric field from $e_0 = 0.05$ to $e_0 = 0.2$ strongly accelerates the DW motion in a non-linear fashion.
From the slopes of the DW position at long times we extract the saturation DW velocity $v$. As shown in the inset of Fig. \ref{F5}(a), $v$ scales with the fourth power of the field amplitude as predicted by Eq. \eqref{dotX}. Figure \ref{F5}(b) examines the effect of the driving frequency at a fixed field amplitude $e_0 = 0.1$. Here, the DW velocity increases with frequency from $f = 8$ to $10$ THz, consistent with the enhanced driving force (dots) in Fig.~\ref{F3}(d).

{\it \color{blue}Discussions and Conclusions}. Our model focuses on planar DWs in bulk ferroelectrics. It can be extended to thin films by taking into account the dipolar interaction corrections \cite{Zhou2023,Rodriguez2024}. It would allow us to assess studies that report ballistic ferrons in van der Waals ferroelectrics excited by focused laser beams via impulsive Raman scattering  \cite{Choe2025,Zhang2025} that propagate over large distances with hypersonic velocities \cite{Choe2025}. These platforms provide natural testbeds for our predictions. The position of ferroelectric DWs can be detected by several complementary techniques, including piezoresponse force microscopy \cite{Gruverman2019}, transmission electron microscopy \cite{Jia2008,Nelson2011}, and nonlinear optical microscopy \cite{Lu2018}. Because the ferron-induced momentum flux couples directly to the polarization, the attainable wall velocities are should be of the order of $\sim$km/s, comparable to the highest DW velocities reported in magnetic materials, see e.g. Ref. \cite{Caretta2024}.

Ferrons can also be generated by temperature gradients \cite{Bauer2021,Tang2022ferron}  and/or electrical injections \cite{Shen2025}, where they typically propagate diffusely. Quantitative modeling then requires solving a non-linear Boltzmann equation for disordered ferroelectric textures, which is beyond the scope of the present paper. 

To summarize, we model ferroelectric DW dynamics under ferron currents. In the linear regime, ferrons propagate through walls without reflection. In the nonlinear regime, negative radiation pressure provides a drag that pulls the wall back towards the source.  The effect applies to move an expanding zoo of topological ferroelectric textures, including other (90-degree, Bloch, Néel) types of DWs,  bubbles \cite{Aramberri2024,Gonzalez2024}, merons \cite{Wang2020}, vortices \cite{Nelson2011NL,Yadav2016,Abid2021}, and skyrmions \cite{Nahas2015,Das2019,Goncalves2019}. Our results open new pathways to rapidly and energy-efficiently control ferroelectric DWs and related topological textures.


\begin{acknowledgments}
We would like to thank Ping Tang and Zhaozhuo Zeng for helpful discussions and acknowledge support from the Center for Science and Innovation in Spintronics (CSIS), Tohoku University. This work was funded by JSPS KAKENHI Grants No. 22H04965 and No. 24H02231. P. Y. acknowledges the financial support from the National Key R\&D Program of China (Grants No. 2025YFA1411302 and No. 2022YFA1402802), the National Natural Science Foundation of China (NSFC) (Grants No. 12374103 and No. 12434003), and Sichuan Science and Technology Program (Grant No. 2025NSFJQ0045).
\end{acknowledgments}


\begin{thebibliography}{99}
\bibitem{PTpotential} The standard P\"{o}schl-Teller potential is expressed as $V(x)=-\lambda(\lambda+1){\rm sech}^2(x)$. In our case the index is the integer $\lambda=2$, so the spectrum contains two bound states and a reflectionless scattering continuum.
\end{thebibliography}
\end{document}